\documentclass{article}

\usepackage[nonatbib,final]{nips_2017}

\usepackage[utf8]{inputenc} 
\usepackage[T1]{fontenc}    
\usepackage{hyperref}       
\usepackage{url}            
\usepackage{booktabs}       
\usepackage{amsfonts}       
\usepackage{nicefrac}       
\usepackage{microtype}      
\usepackage{graphicx}
\usepackage{soul,color}

\newcommand{\Lagr}{\mathcal{L}}

\title{A Wavenet for Speech Denoising}

\author{
  Dario Rethage\thanks{Contributed equally.} \\  \texttt{dario@rethage.net} \\ Music Technology Group\\ Universitat Pompeu Fabra	\And
  Jordi Pons\footnotemark[1] \\ \texttt{jordi.pons@upf.edu} \\ Music Technology Group\\ Universitat Pompeu Fabra
  \And
  Xavier Serra \\ \texttt{xavier.serra@upf.edu} \\ Music Technology Group\\ Universitat Pompeu Fabra
}

\begin{document}

\maketitle
\vspace{-5mm}
\begin{abstract}
Currently, most speech processing techniques use magnitude spectrograms as front-end and are therefore by default discarding part of the signal: the phase. In order to overcome this limitation, we propose an end-to-end learning method for speech denoising based on Wavenet. The proposed model adaptation retains Wavenet's powerful acoustic modeling capabilities, while significantly reducing its time-complexity by eliminating its autoregressive nature. Specifically, the model makes use of non-causal, dilated convolutions and predicts target fields instead of a single target sample. The discriminative adaptation of the model we propose, learns in a supervised fashion via minimizing a regression loss. These modifications make the model highly parallelizable during both training and inference. 
Both computational and perceptual evaluations indicate that the proposed method is preferred to Wiener filtering, a common method based on processing the magnitude spectrogram.

\end{abstract}

\section{Introduction}

Over the last several decades, machine learning has produced solutions to complex problems that were previously unattainable with signal processing techniques \cite{chen2017deep,krizhevsky2012imagenet,zhang2016stackgan}.
Speech recognition is one such problem where machine learning has had a very strong impact. However, until today it has been standard practice not to work directly in the time-domain, but rather to explicitly use time-frequency representations as input \cite{amodei2016deep, weninger2015speech,xiong2016microsoft} -- for reducing the high-dimensionality of raw waveforms. Similarly, most techniques for speech denoising use magnitude spectrograms as front-end 
\cite{kumar2016speech,lu2013speech,parveen2004speech, weninger2015speech,xu2015regression}.
Nevertheless, this practice comes with its drawbacks of discarding potentially valuable information (phase) and utilizing general-purpose feature extractors (magnitude spectrogram analysis) instead of learning specific feature representations for a given data distribution. 

Most recently, neural networks have shown to be effective in handling structured temporal dependencies between samples of a discretized audio signal. For example, consider the most local structure of a speech waveform ($\approx$ tens of milliseconds). In this range of context, many sonic characteristics of the speaker (timbre) can be captured and linguistic patterns in the speech become accessible in the form of phonemes. It is important to note that these levels of structure are not discrete, making techniques that explicitly focus on different levels of structure inherently suboptimal. This suggests that deep learning methods, capable of learning multi-scale structure directly from raw audio, may have great potential in learning such structures. To this end, discriminative models have been used in a end-to-end learning fashion for music \cite{dieleman2014end,lee2017sample} or speech classification \cite{collobert2016wav2letter,palaz2015convolutional,zhu2016learning}. Raw audio waveforms have also been successfully employed for generative tasks \cite{engel2017neural,mehri2016samplernn,van2016wavenet,pascual2017segan}. Interestingly, most of these generative models are autoregressive \cite{engel2017neural,mehri2016samplernn,van2016wavenet}, with the exception of SEGAN -- which is based on a generative adversarial network \cite{pascual2017segan}. We are not aware of any generative model for raw audio based on variational autoencoders.

Previous discussion motivates our study in adapting Wavenet's model (an autoregressive generative model) for speech denoising. Our main hypothesis is that by learning multi-scale hierarchical representations from raw audio we can overcome the inherent limitations of using the magnitude spectrogram as a front-end for this task. 
Some work in this direction already exists. Back in the 80s, Tamura et al. \cite{tamura1988noise} used a four-layered feed-forward network operating directly in the raw-audio domain to learn a noise-reduction mapping. And recently: Pascual et al. \cite{pascual2017segan} proposed the use of an end-to-end generative adversarial network for speech denoising, and Qian et al. \cite{BaWN} used a Bayesian Wavenet for speech denoising. In all three cases, they provide better results than their counterparts based on processing magnitude spectrograms.

Section 2 describes the original Wavenet architecture, and section 3 describes the modifications we propose. In section 4 we experiment with and discuss some architectural parameters. And finally, section 5 concludes by highlighting the most relevant contributions.

\vspace{-2mm}
\section{Wavenet}
\vspace{-2mm}

\begin{figure}
\centering
  \includegraphics[width=0.9\textwidth]{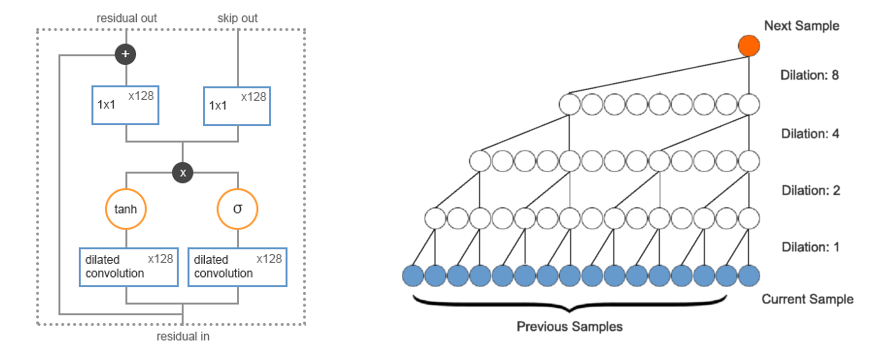}
  \caption{\textit{Left} -- Residual layer. \textit{Right} -- Causal, dilated convolutions with increasing dilation factors.}
  \label{fig:wavenet_dilated_layers}
\end{figure}

Wavenet is capable of synthesizing natural sounding speech \cite{van2016wavenet}. This autoregressive model shapes the probability distribution of the next sample given some fragment of previous samples. The next sample is produced by sampling from this distribution. An entire sequence of samples is produced by sequentially feeding previously generated samples back into the model -- this enforces time continuity in the resulting audio waveforms. A high-level visual depiction of the model is presented in Figure~\ref{fig:overview_wavenet}. Wavenet is the audio domain adaptation of the PixelCNN generative model for images \cite{oord2016pixel,van2016conditional}. Wavenet retains many PixelCNN features like: causality, gated convolutional units, discrete softmax output distributions and the possibility of conditioning the model -- while introducing dilated convolutions and non-linear quantization \cite{van2016wavenet}. Some of Wavenet's key features are presented below:
\vspace{-2mm}
\begin{figure}[h]
\centering
  \includegraphics[width=0.7\textwidth]{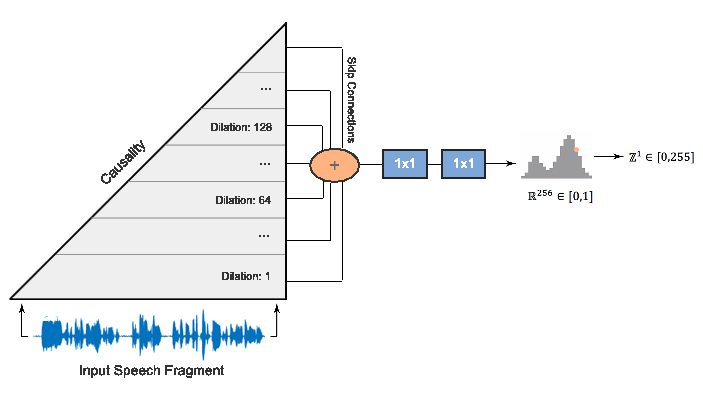}
  \caption{Overview of Wavenet.}
  \label{fig:overview_wavenet}
\end{figure}
\vspace{-2mm}
\paragraph{Gated Units} As in LSTMs \cite{hochreiter1997long}, sigmoidal gates control the activations' contribution in every layer:
$z_{t'} = \tanh(W_{f} * x_{t}) \odot \sigma(W_{g} * x_{t})$,
where $*$ and $\odot$ operators denote convolution and element-wise multiplication, respectively. $f$, $t$, $t'$ and $g$ stand for filter, input time, output time and gate indices. $W_f$ and $W_g$ are convolutional filters. Figure~\ref{fig:wavenet_dilated_layers} (\textit{Left}) depicts how sigmoidal gates are utilized.

\paragraph{Causal, dilated convolutions}

Wavenet makes use of causal, dilated convolutions \cite{van2016wavenet,yu2015multi}. It uses a series of small (length = 2) convolutional filters with exponentially increasing dilation factors. This results in a exponential receptive field growth with depth. Causality is enforced by asymmetric padding proportional to the dilation factor, which prevents activations from propagating back in time -- see Figure~\ref{fig:wavenet_dilated_layers} (\textit{Right}).
Each dilated convolution is contained in a residual layer \cite{he2016deep}, controlled by a sigmoidal gate with an additional 1x1 convolution and a residual connection -- see Figure~\ref{fig:wavenet_dilated_layers} (\textit{Left}).

\paragraph{$\mu$-law quantization}

When using a discrete softmax output distribution, it is necessary to perform a more coarse 8-bit quantization to make the task computationally tractable. This is accomplished via a $\mu$-law non-linear companding followed by an 8-bit quantization (256 possible values):
\[f(x_{t}) = sign(x_{t})\frac{\ln(1+\mu|x_{t}|)}{\ln(1 + \mu)}\]
\paragraph{Skip Connections}

These offer two advantages. First, they facilitate training deep models \cite{szegedy2015going}. And second, they enable information at each layer to be propagated directly to the final layers. This allows the network to explicitly incorporate features extracted at several hierarchical levels into its final prediction \cite{lee2017multi}. Figure~\ref{fig:overview_wavenet} and \ref{fig:wavenet_dilated_layers} (\textit{Left}) provide further details in how skip connections are used.

\paragraph{Context Stacks}

These deepen the network without increasing the receptive field length as drastically as increasing the dilation factor does. This is achieved by simply stacking a set of layers, dilated to some maximum dilation factor, onto each other -- and can be done as many times as desired \cite{van2016wavenet}. For example, Figure~\ref{fig:wavenet_dilated_layers} (\textit{Right}) is composed of a single stack.

\paragraph{Time-complexity}

A significant drawback of Wavenet is its sequential (non-parallelizable) generation of samples. This limitation is strongly considered in the speech-denoising Wavenet design.

\section{Wavenet for Speech Denoising}

Speech denoising techniques aim to improve the intelligibility and the overall perceptual quality of speech signals with intrusive background-noise. The problem is typically formulated as follows: $m_{t} = s_{t} + b_{t}$, 
where: $m_{t}$ $\equiv$ mixed signal, $s_{t}$ $\equiv$ speech signal, $b_{t}$ $\equiv$ background-noise signal. The goal is to estimate $s_{t}$ given $m_{t}$. Speech denoising, while sharing many properties with speech synthesis, also has several unique characteristics -- these motivated the design of this Wavenet adaptation. A high-level visual depiction of the model is presented in Figure~\ref{fig:gen_ss_overview}. Its key features are presented below:
\begin{figure}[!h]
\centering
  \includegraphics[width=0.7\textwidth]{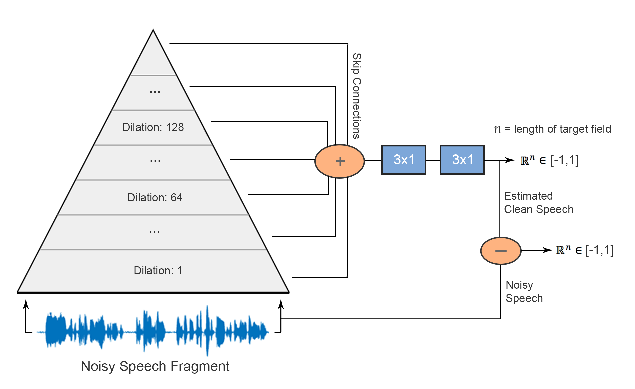}
  \caption{Overview of the speech-denoising Wavenet.}
  \label{fig:gen_ss_overview}
\end{figure}

\begin{figure}
\centering
  \includegraphics[width=0.85\textwidth]{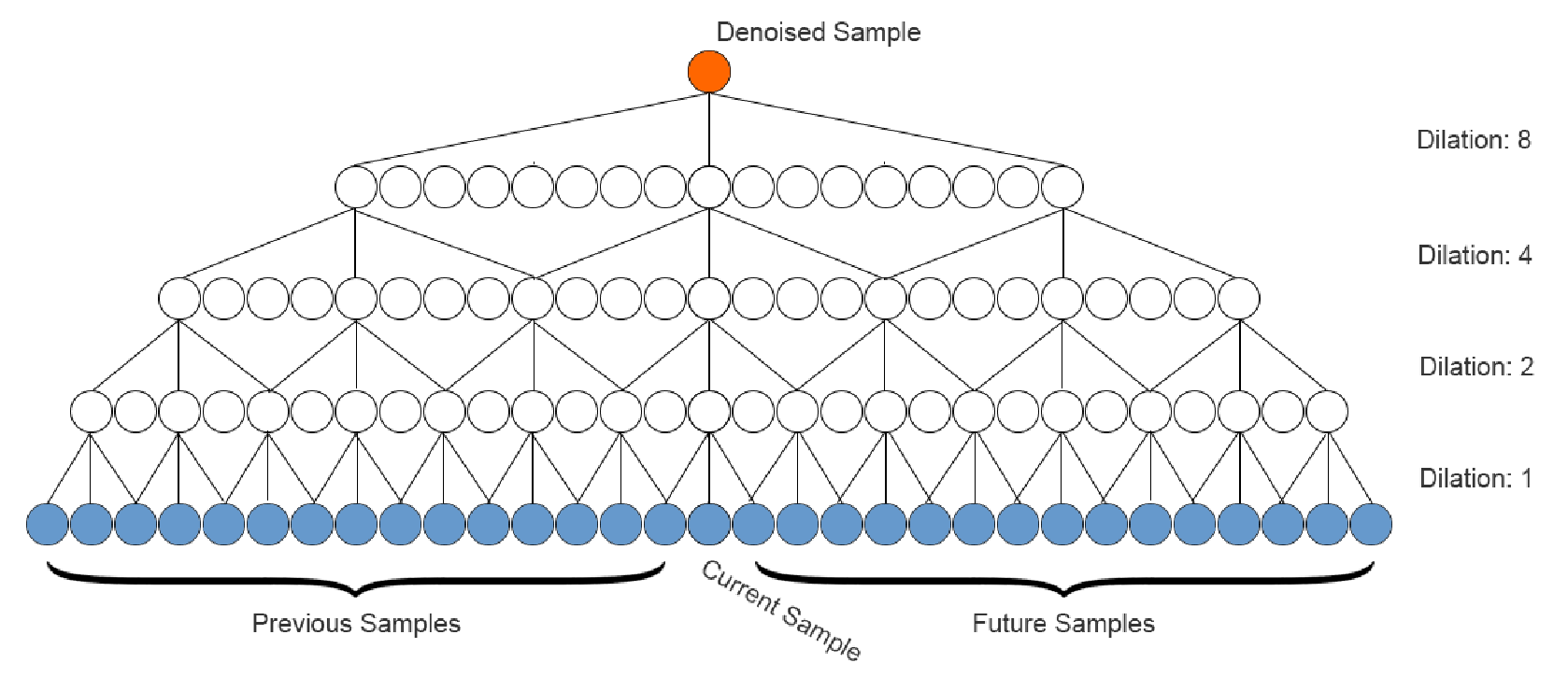}
  \caption{Non-causal, dilated convolutions with exponentially increasing dilation factors.}
  \label{fig:gen_ss_dilated_layers}
\end{figure}

\paragraph{Non-causality}
Contrary to audio synthesis, in speech denoising, some future samples are generally available to help make more well informed predictions.
Even in real time applications, when a few milliseconds of latency in model response can be afforded, the model has access to valuable information about samples occurring shortly after a particular sample of interest. As a result, and given that Wavenet's time-complexity was a major constraint, the autoregressive causal nature of it was removed in the proposed model.
A logical extension to Wavenet's asymmetric dilated convolution pattern (shown in Figure~\ref{fig:wavenet_dilated_layers}) is to increase the filter length to 3 and perform symmetric padding at each dilated layer. If the sample we wish to enhance is now taken to be at the center of the receptive field, this has the effect of doubling the context around a sample of interest and eliminating causality.
The model has access to the same amount of samples in the past as samples in the future to inform the prediction. This can be seen in Figure~\ref{fig:gen_ss_dilated_layers}. Early experiments with larger filters of length 5, 11 and 21 showed inferior performance.

\begin{figure}
\centering
  \includegraphics[width=0.85\textwidth]{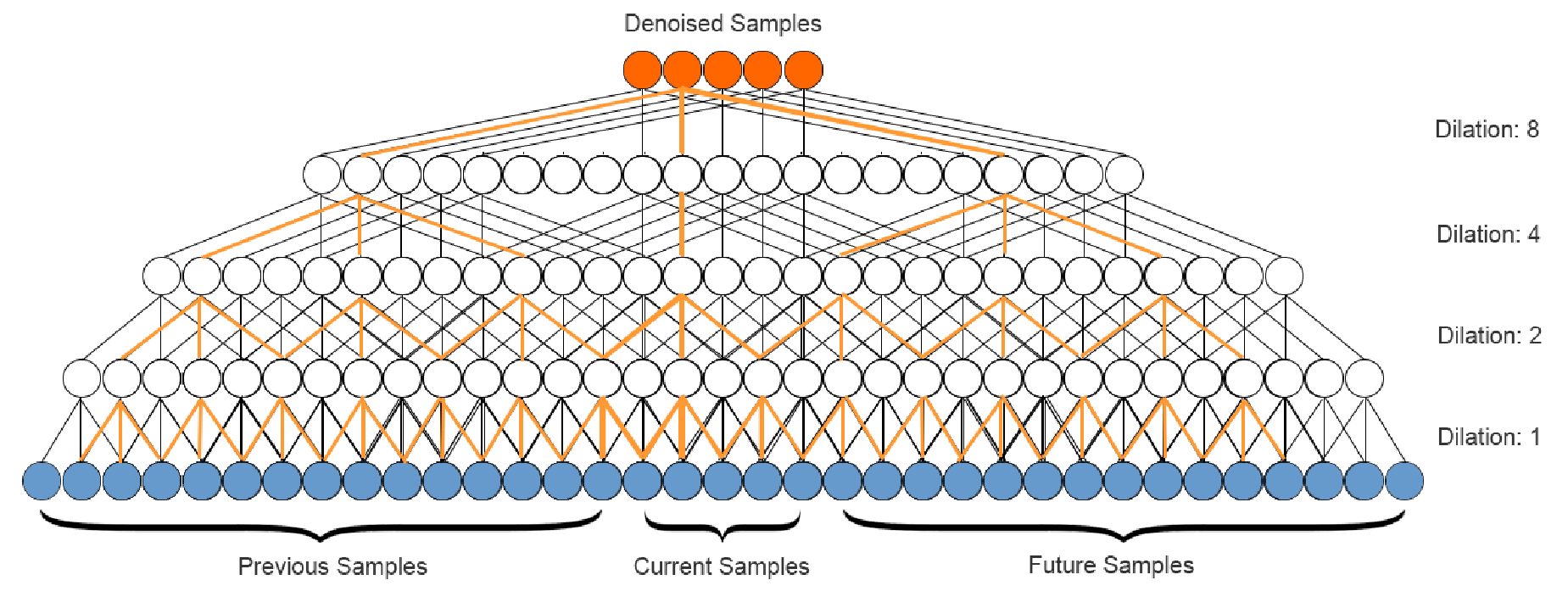}
  \caption{Predicting on a target field.}
  \label{fig:gen_ss_target_field_dilated_layers}
\end{figure}

\paragraph{Real-valued predictions}
Wavenet uses a discrete softmax output to avoid making any assumption on the shape of the output's distribution -- this is suitable for modeling multi-modal distributions. However, early experiments with discrete softmax outputs proved disadvantageous. Instead, the potentially multi-modal output distribution allowed for artifacts to be introduced into the denoised signal. This suggests that real-valued predictions (assuming uni-modal gaussian-shaped output distributions) seem to be more appropriate for our problem.
Moreover, experiments with discrete softmax outputs resulted in output distributions with high variance -- signifying low confidence with the value having highest probability. $\mu$-law quantization was also disadvantageous because it disproportionately amplified the background-noise. For these reasons, the proposed model predicts raw audio -- without any pre-processing.

\clearpage
\paragraph{Energy-conserving loss}
We propose utilizing a two term loss that enforces energy conservation:
\[ \Lagr(\hat{s}_{t}) = |s_{t}-\hat{s}_{t}| + |b_{t}-\hat{b}_{t}|\]

The background-noise signal is estimated by subtracting the denoised speech from the mixed input, as shown in Figure~\ref{fig:gen_ss_overview} -- and in the following formulation: $\hat{b}_{t} = m_{t} - \hat{s}_{t}$. 
As a result, the network predicts both components of the signal and, since the second output is produced by a parameterless operation on the speech estimate, the loss produced by this output is directly representative of the model's ability to perform the actual task of interest. Note that: $\hat{m_{t}} = \hat{s}_{t} + \hat{b}_{t} = \hat{s}_{t} + (m_{t} - \hat{s}_{t}) = m_{t} \rightarrow \hat{m}_{t} =  m_{t}$, \textrm{then:} $\hat{s}_{t} + \hat{b}_{t} =  s_{t} + b_{t} \rightarrow 0 = s_{t} - \hat{s}_{t} + b_{t} - \hat{b}_{t}$. Therefore, the proposed way of estimating $\hat{b}_{t}$ tailors the two term loss towards conserving the energy of the original mixture: $E_{\hat{m}_{t}} \equiv E_{m_{t}}$.
Previous work considered energy-conservative pipelines: for source separation  \cite{pons2016remixing,chandna2017monoaural}, or for speech enhancement \cite{weninger2015speech}. 
Finally, note that the first-term of the proposed loss corresponds to the standard L1 loss -- that is used as baseline for comparison in Table~\ref{computed-mos-measures}.

\paragraph{Discriminative model} 

Note that the proposed model is no longer autoregressive and its output is not explicitly modeling a probability distribution, but rather the output itself. Furthermore, the model is trained in a supervised fashion -- by minimizing a regression loss function. As a result: the proposed model is no longer generative (like Wavenet), but discriminative.

\paragraph{Final 3x1 filters}

As mentioned, the proposed architecture is not autoregressive. That is, previously generated samples are not fed back into the model to inform future predictions -- which enforces time continuity in the resulting signal. Early experiments produced waveforms with sporadic point discontinuities that sounded very disruptive. Replacing the kernels of the two final layers with 3x1 filters instead of 1x1 filters reimposed this constraint. Larger kernels were also considered, although these did not improve our results.

\paragraph{Target field prediction}

The proposed model does not predict just one, but a set of samples in a single forward propagation -- see Figure~\ref{fig:gen_ss_target_field_dilated_layers}.
Parallelizing the inference process from 1 sample to on the order of 1000 samples offers significant memory and time savings. This is because overlapping data is used for predicting neighboring samples, and by predicting target fields these redundant computations are done just once. 

The receptive field length ($rf$) of the model is the number of input samples that go into the prediction of a single denoised output sample. In order to maintain that every output sample in the target field ($tf$) has a full receptive field of context contributing to its prediction, the length of the fragment presented to the model must be equal to: $rf + (tf - 1)$. Finally, note that the cost is computed sample-wise -- during training, individual sample costs of a target field are averaged.

\paragraph{Conditioning} The model is conditioned on a binary-encoded scalar corresponding to the identity of the speaker. This condition value is the bias term in every convolution operation. Condition values 1-28 represent each of the 28 speakers comprising the training set. 
In addition, we add an auxiliary code (all zeros) denoting any speaker identity so that the trained model can be used for unknown speakers. The same training data is presented to the model either conditioned to its speaker ID or to zeros. 1/29\% of training samples have their condition values set to zeros. Note that this conditioning procedure can also be interpreted as a data augmentation strategy.

\paragraph{Noise-only data augmentation}

A form of augmentation in which a proportion of training samples contain only background-noise was also employed after observing that our model had difficulties producing silence. In section~\ref{objective-evaluation} we experiment with 10\% and 20\% noise-only training samples.

\paragraph{Denoising step} The network is presented with a noisy speech fragment and the condition value is set to zero. By default the network denoises the input in batches, iteratively appending each denoised fragment to the previous.
Alternatively, the fully-convolutional architecture we use is flexible in the time domain. Therefore, it permits denoising on a different length of audio than the one used during training. This allows the model to denoise an entire audio sample in one-shot -- given sufficient memory availability\footnote{On a Titan X Pascal (12GB-VRAM) it was possible to denoise up to 25s of audio using one-shot denoising.}.

\section{Experiments}

\vspace{-1mm}

\subsection{Dataset}

\vspace{-1mm}

The used dataset \cite{pascual2017segan,valentiniinvestigating} was generated from two sources: speech data was supplied by the Voice Bank corpus \cite{veaux2013voice} while environmental sounds were provided by the Diverse Environments Multichannel Acoustic Noise Database (DEMAND) \cite{thiemann2013diverse}. The subset of the Voice Bank corpus we used features 30 native english speakers from different parts of the world reading out $\approx$ 400 sentences -- 28 speakers are used for training and 2 for testing. Recordings are of studio quality sampled at 48kHz -- and subsampled to 16kHz for this study. The subset of DEMAND that we used provides recordings in 13 different environmental conditions such as in a park, in a bus or in a cafe -- 8 background-noises are mixed with speech during training and 5 background-noises are used during testing. DEMAND was produced with a 16-channel array sampled at 48kHz, however for the purposes of this work all channels were merged and subsampled to 16kHz. During training, two artificial noise classes were added -- in total 10 different noise classes are available during training. Training samples are synthetically mixed at one of the following four signal-to-noise ratios (SNRs): 0, 5, 10 and 15dB with one of the 10 noise types. This results in 11,572 training samples from 28 speakers under 40 different noise conditions. Test samples are also synthetically mixed at one of the following four different SNRs: 2.5, 7.5, 12.5 and 17.5dB with one of the 5 test-noise types -- resulting in 20 noise conditions for 2 speakers. As a result, the test set features 824 samples from unseen speakers and noise conditions. For both sets, the samples are on average 3 seconds long with a standard deviation of 1 second. No preprocessing to the audio (such as pre-emphasis filtering \cite{pascual2017segan} or $\mu$-law quantization\cite{van2016wavenet}) is used, allowing the pipeline to be end-to-end in the strictest sense. 

\vspace{-1mm}

\subsection{Basic experimental setup}

\vspace{-1mm}

The proposed model features 30 residual layers -- as in Figure~\ref{fig:wavenet_dilated_layers} (\textit{Left}). The dilation factor in each layer increases in the range 1, 2, ..., 256, 512 by powers of 2. This pattern is repeated 3 times (3~stacks). Prior to the first dilated convolution, the 1-channel input is linearly projected to 128 channels by a standard 3x1 convolution to comply with the number of filters in each residual layer. The skip connections are 1x1 convolutions also featuring 128 filters -- a RELU is applied after summing all skip connections. The final two 3x1 convolutional layers are not dilated, contain 2048 and 256 filters respectively, and are separated by a RELU. The output layer linearly projects the feature map into a single-channel temporal signal by using a 1x1 filter. This parameterization results in a receptive field of 6,139 samples ($\approx$ 384ms). The target field is comprised of 1601 samples ($\approx$ 100ms) -- optimized to adhere our memory constraints. The relatively small size of the model (6.3 million parameters) together with its parallel inference on 1601 samples at once, results in a denoising time of $\approx$ 0.56 seconds per second of noisy audio on GPU. Unless explicitly stated, we assume using the following basic experimental setup: training was done with the energy-conserving loss, conditioning to speaker ID and without data augmentation. Code and trained models are available online\footnote{https://github.com/drethage/speech-denoising-wavenet}.

We set as baselines for comparison: \textit{i)} the noisy signal, and \textit{ii)} a signal processing method based on Wiener filtering -- a widely used technique for speech-denoising \cite{weninger2015speech,scalart1996speech} or source-separation \cite{carabias2013nonnegative,chandna2017monoaural,pons2016remixing,huang2015joint}. The baseline algorithm uses a Wiener filtering method based on a priori SNR estimation \cite{scalart1996speech}, as~implemented here\footnote{\url{https://www.crcpress.com/downloads/K14513/K14513\_CD\_Files.zip}}.

\vspace{-1mm}

\subsection{Evaluation based on computational measurements}
\label{objective-evaluation}

\vspace{-1mm}
  
The goal is to measure the quality of the denoised speech along three dimensions: signal distortion, background-noise interference and overall quality. To this end, we consider three measures \cite{hu2006evaluation}: \textbf{SIG}~-~predictor of signal distortion; \textbf{BAK} - background-noise intrusiveness predictor; and
\textbf{OVL}~-~predictor of overall quality. These measures operate in a 1--5 range, aiming to computationally approximate the Mean Opinion Score (MOS) that would be produced from human perceptual trials.

Experiments were conducted to study how noise-only data augmentation, target field length, energy-conserving loss and conditioning influence the model's performance.
Computed MOS measures relating to these experiments are presented together in Table \ref{computed-mos-measures}. Italicized rows correspond to the basic experimental setup introduced above. Non-italicized rows represent a single parameter modification of this basic setup.

\begin{table}[h]
  \caption{Computed MOS measures on test set. Ranging from 1--5, higher scores are better.}
  \label{computed-mos-measures}
  \centering
\def\arraystretch{1.3}
 \begin{tabular} { l c c c c l c c c }
 \toprule
 \textbf{Model} & \textbf{SIG} & \textbf{BAK} &  \textbf{OVL} & \hspace{7mm} & \textbf{Model} & \textbf{SIG} & \textbf{BAK} & \textbf{OVL}\\
 \toprule
 \multicolumn{4}{c}{\textbf{Noise-only data augmentation}} &  & \multicolumn{4}{c}{\textbf{Target field length}} \\
 20\% & 2.74 & 2.98 & 2.30 & & 1 sample* & 1.37 & 1.79 & 1.28\\
 10\% & 2.95 & 3.12 & 2.49 &  & 101 samples* & 1.67 & 2.07 & 1.50\\
 \textit{0 \%}& \textit{3.62} & \textit{3.23} & \textit{2.98} & & \textit{1601 samples}& \textit{3.62} & \textit{3.23} & \textit{2.98}\\
 \hline
 \multicolumn{4}{c}{\textbf{Loss}} & &  \multicolumn{4}{c}{\textbf{Conditioning}}  \\
 L1 & 3.54 & 3.22 & 2.93 &  &  Unconditioned & 3.48 & 3.12 & 2.88\\
 \textit{Energy-Conserving} & \textit{3.62} & \textit{3.23} & \textit{2.98} &  & \textit{Conditioned} & \textit{3.62} & \textit{3.23} & \textit{2.98}\\ \toprule
 \textbf{Wiener filtering} & 3.52 & 2.93 & 2.90 &  & \textbf{Noisy signal} & 3.51 & 2.66 & 2.79 \\
 \bottomrule
 \end{tabular}
*Computed on perceptual test set due to computational (time) constraints.
\end{table}

\vphantom{}
\vspace{-9mm}
The basic experimental setup with no data augmentation (0 \%) achieves better results across all metrics. However, informal listening clearly shows that training with 10\% noise-only augmentation allows the model to produce silence in moments where no speech is present without degrading the signal, which is perceptually pleasant when aurally evaluating denoised samples. 20 \% noise-only augmentation achieves lower computed MOS ratings and further listening reveals that it does not improve the quality of the denoised samples.

Moreover, we observe that training with longer target fields is crucial in achieving any significant denoising. Models trained with small target field lengths produce audio which not only fails to denoise, but also sounds fuzzier than the input. In addition, we observe that models with a small target field length require impractically long inference times (as a result of many redundant computations). 
Due to this, the results for smaller target field lengths in Table~\ref{computed-mos-measures} are computed with the 20-sample perceptual test set.

Although the improvement is barely noticeable aurally, the proposed energy-conserving loss also achieves slightly better results than the standard L1 loss. This might be occurring because the energy conserving loss is simply two times the L1 loss -- what is caused by the way the proposed model computes the background-noise estimate\footnote{Since the background-noise estimate is computed by a parameterless subtraction ($\hat{b}_{t} = m_{t} - \hat{s}_{t}$), then: $\Lagr(\hat{s}_{t}) = |s_{t}-\hat{s}_{t}| + |b_{t}-\hat{b}_{t}| = L_1(\hat{s}_{t}) + |(m_{t} - s_{t})-(m_{t} - \hat{s}_{t})| = L_1(\hat{s}_{t}) + |- s_{t} + \hat{s}_{t}| = 2\cdot L_1(\hat{s}_{t})$}. We leave for future work studying the impact of the proposed loss in cases where the background-noise is estimated through a more powerful model.


Results suggest that conditioning on speaker ID is beneficial for achieving a better speech denoising. Improvements are consistent throughout measures, although these are marginal. Informal listening confirms the results depicted by the computational measurements: the background-noise and algorithmic artifacts in conditioned samples are marginally, but noticeably reduced.

When comparing the proposed model with the baseline Wiener filtering method, one observes that OVL and SIG results are comparable, showing that Wiener filtering similarly preserves the quality of the speech signal. However, the proposed method removes the background-noise more effectively than Wiener filtering. In line with this, informal listening confirms that Wiener filtering takes small risks: no strong speech signal distortion is measured, but likewise background-noise is not heavily removed. When comparing the Wiener filtering MOS measures with the ones measured on the noisy signal, one observes that the baseline algorithm denoises without introducing algorithmic artifacts.

As seen, computational measures alone do not reveal a clear best performing configuration. Based on previously discussed informal listening and computed MOS measures, we consider the following model to achieve the best sounding results: conditioned, energy-conserving loss, with a target field of 1601 samples and 10\% noise-only data augmentation. 
\vspace{-1mm}
\subsection{Perceptual evaluation}
\vspace{-1mm}
Perceptual tests were conducted with 33 participants to get subjective feedback on the effectiveness of the speech-denoising Wavenet. 20 audio samples were chosen to compose the perceptual test set: 5 samples from each of the four SNRs, with an equal number of samples coming from each of the 2 speakers in the test set. Aside from these constraints, the samples were chosen randomly. Participants were presented with 4 variants of each sample: \textit{i)} the original mix with speech and background-noise, \textit{ii)} clean speech, \textit{iii)} speech denoised by Wiener filtering, and \textit{iv)} speech denoised with the best performing Wavenet -- defined at the end of section~\ref{objective-evaluation}. Participants were asked to \textit{``give an overall quality score, taking into consideration both: speech quality and background-noise suppression"} for each of the last two variants. The first two variants were presented as references. Participants were able to give a score between 1--5, with a 1 being described as \textit{``degraded speech with very intrusive background"} and a 5 being \textit{``not degraded speech with unnoticeable background"} \cite{hu2006evaluation,loizou2011speech}. MOS quality measurement is obtained by averaging the scores from all participants. We also compute the t-test ($H_0$ being that means are equal) to study whether obtained results are statistically significant or not.
Table \ref{subjective-mos-measures} presents the results of the perceptual evaluation, showing that participants significantly preferred (t-test: \textit{p}-value < 0.001) the proposed method over the one based on Wiener filtering.
\vspace{-1mm}
\begin{table}[h]
  \caption{Subjective MOS measures on perceptual test set. Ranging from 1--5, higher scores are better.}
  \label{subjective-mos-measures}
  \centering
\def\arraystretch{1.3}
 \begin{tabular} { c c c}
 \toprule
 \textbf{Measurement} & \textbf{Wiener filtering} & \textbf{Proposed Wavenet} \\
 \toprule
 MOS & 2.92 & 3.60\\ 
 \bottomrule
 \end{tabular}
\end{table}
\vspace{-8mm}
\section{Conclusion}
\vspace{-2mm}
We have presented a discriminative adaptation of Wavenet's model for speech denoising that features a non-causal and non-autoregressive architecture. This allows us to reduce the time-complexity of the model, one of the main drawbacks of Wavenet.  However, removing autoregression also means that temporal continuity in the resulting signal is no longer enforced. In order to overcome this limitation, we observe that adjacent continuity is maintained by using 3x1 filters in the final layers of the model. 

The proposed model is able to predict target fields instead of single samples -- which further reduces its time-complexity while also significantly improving the performance of the model. In addition, the convolutional nature of the model makes it flexible in the time-dimension.  As a result, it supports denoising variable-length audio -- independently of how the model was trained. Therefore, the proposed model allows for one-shot denoising. This flexibility can be valuable since training and inference may be done on different hardware with varying memory availability.

Algorithmic artifacts appeared early on in our research. Switching to real-valued outputs was key in removing these artifacts introduced by the softmax loss used in Wavenet. Interestingly, note that directly operating in the raw audio domain enables considering alternative costs that can be motivated from a domain knowledge perspective. 

Initially, it was also challenging for the model to generate silence. To mitigate this, we use the proposed noise-only data augmentation which has shown to be effective for this. However, the model still has its limitations, \textit{i.e.:} its inability to deal with sudden interferences like honks in city traffic.

Although our focus is speech denoising, it is worth noting that the proposed model inherently estimates two sources: speech and background-noise -- since the background-noise can be computed via subtracting the speech estimate to the input. These results are in line with recent work \cite{venkataramani2017end} showing that end-to-end pipelines are starting to prove effective for source separation tasks, as~well.

It is important to note that no speech specific constraints are incorporated into our pipeline to overcome the aforementioned challenges. Instead, we either propose architectural improvements to our model or we propose new forms of data augmentation. The proposed model effectively denoises speech signals under noise conditions and speakers that it has never been exposed to -- audio samples are available online for listening\footnote{Speech and background-noise estimates: \url{http://jordipons.me/apps/speech-denoising-wavenet}}. This implies that the proposed discriminative reformulation of Wavenet does not sacrifice its modeling capabilities while significantly reducing its time-complexity. 

Further, our adaptation of Wavenet for speech denoising differs from the Bayesian variant \cite{BaWN} in many ways, particularly that: \textit{i)} we do not explicitly manipulate probability distributions, and \textit{ii)} that our model does not infer sequentially.

Finally, perceptual tests show that our model's estimates are preferred over the ones based on Wiener filtering. This confirms that it is possible to learn multi-scale hierarchical representations from raw audio instead of using magnitude spectrograms as front-end for the task of speech denoising.

\vspace{-2mm}
\section{Acknowledgments}
\vspace{-2mm}
This work is partially supported by the Maria de Maeztu Units of Excellence Programme (MDM-2015-0502). We are grateful for the GPUs donated by NVidia, and also thanks to \url{http://foxnice.com} for hosting our demos. 

\bibliographystyle{plain}
\bibliography{references}

\end{document}